\documentclass{aa}  

\usepackage{natbib}
\bibpunct{(}{)}{;}{a}{}{,} 

\usepackage{graphicx}
\usepackage{tikz}
\usetikzlibrary{arrows,shapes,automata,backgrounds,petri,positioning}

\usepackage{amssymb,amsmath,bm}
\usepackage{delarray}
\usepackage{mathrsfs}
\usepackage[cal=boondox]{mathalfa}
\usepackage{xcolor}
\usepackage{soul}
\usepackage[varg]{txfonts}
\usepackage{hyperref}
\hypersetup{
  colorlinks   = true, 
  urlcolor     = blue, 
  linkcolor    = blue, 
  citecolor   = blue 
}

\newcommand{\ff}{{\mathbf f}}
\newcommand{\rr}{{\mathbf r}}
\newcommand{\xx}{{\mathbf x}}
\newcommand{\bb}{{\mathbf b}}

\begin{document} 



\title{Numerical solutions to linear transfer problems\\ 
of polarized radiation}
\subtitle{I. Algebraic formulation and stationary iterative methods}

\author{Gioele Janett\inst{1,2}
        \and
        Pietro Benedusi\inst{2}
        \and
        Luca Belluzzi\inst{1,3,2}
        \and
        Rolf Krause\inst{2}}
\institute{
Istituto Ricerche Solari (IRSOL), Universit\`a della Svizzera italiana (USI), CH-6605 Locarno-Monti, Switzerland 
          \and
Euler Institute, Universit\`a della Svizzera italiana (USI), CH-6900 Lugano, Switzerland
          \and
Leibniz-Institut f\"ur Sonnenphysik (KIS), D-79104 Freiburg i.~Br., Germany
\\      \email{gioele.janett@irsol.usi.ch}
}

\abstract
{
The numerical modeling of the generation and transfer of polarized radiation is a key task in solar and stellar physics research and has led to a relevant class of discrete problems that can be reframed as linear systems.
In order to solve such problems, it is common to rely on efficient stationary iterative methods.
However, the convergence properties of these methods are problem-dependent, and
a rigorous investigation of their convergence conditions, when applied to transfer problems of polarized radiation, is still lacking.
}
{
After summarizing the most widely employed iterative methods used in the numerical
transfer of polarized radiation, this article aims to clarify
how the convergence of these methods depends on different design elements,
such as the choice of the formal solver, the discretization of the problem, or the use of damping factors.
The main goal is to highlight advantages and
disadvantages of the different iterative methods in terms of stability and rate of convergence.}
{
We first introduce an algebraic formulation of the radiative transfer problem.
This formulation allows us to explicitly assemble the iteration matrices
arising from different stationary iterative methods,  compute their spectral radii and derive their convergence rates, and test the impact of different discretization settings, problem parameters, and damping factors.
}
{
Numerical analysis shows that the choice of the formal solver significantly affects, and can even prevent, the convergence 
of an iterative method. 
Moreover, the use of a suitable damping factor can both enforce stability and increase the convergence rate.
}
{
The general methodology used in this article, based on a fully algebraic formulation of linear 
transfer problems of polarized radiation,
provides useful estimates of the convergence rates of various iterative schemes.
Additionally, it can lead to novel solution approaches as well as analyses for a wider range of settings, including the unpolarized case.
}


\maketitle

\section{Introduction}\label{sec:sec1}
%
The theory of the generation and transfer of polarized radiation is of crucial importance
for solar and stellar physics. Currently, one of the main challenges in the field
is performing detailed numerical simulations of the
transfer of polarized radiation in magnetized and dynamic atmospheres.
In particular, it is essential to properly account for scattering processes in complex
atomic models and multidimensional geometries, which lead to nonlinear and nonlocal multidimensional problems.
In applied science,
it is often profitable to convert nonlinear problems into linear ones, allowing the use of the powerful tools from numerical linear algebra, such as stationary or Krylov iterative methods.
This is also the case in the field of numerical radiative transfer, where a fair number of problems of practical 
interest are accurately described through a linear system.
For instance, the modeling of the transfer of resonance line polarization considering two-level (or two-term) atomic models
can be linearized by assuming 
that the lower level (or term) is not polarized and that its population is known and fixed a priori
\citep[see, e.g.,][]{belluzzi2014,sampoorna2017,alsinaballester2017}.

In principle, linear problems can be solved in a single step with a direct method.
However, this task can be highly inefficient or even unfeasible in practice, especially for large problems.
Therefore, it is common to rely on efficient stationary iterative methods that, combined with suitable parallelization strategies, 
make the whole problem computationally tractable.

\citet{trujillo_bueno1996} first applied the Richardson (a.k.a. fixed point iteration), Jacobi, and block-Jacobi iterative methods 
to the numerical transfer of polarized radiation\footnote{In the context of linear
radiative transfer problems,
the Richardson method corresponds to the well-known $\Lambda$-iteration, while Jacobi and block-Jacobi methods correspond to the so-called accelerated $\Lambda$-iteration (ALI) methods.},
while \citet{trujillo_bueno1999} applied Gauss-Seidel
and successive over-relaxation (SOR) iterative methods to the same problem.
In particular, the block-Jacobi iterative method is still frequently used in
different scattering polarization problems \citep[e.g.,][]{belluzzi2014,alsinaballester2017,sampoorna2017,alsinaballester2018}.
By contrast, Gauss-Seidel and SOR iterative methods
are not of common practice \citep{lambert2016},
because they are nontrivial to parallelize.
Despite of its slow convergence, the Richardson iterative method
is still used in computationally complex transfer problems, where the other iterative methods
are not exploitable \citep[e.g.,][]{delpinoaleman2020,janett2021b}.

A rigorous investigation of the convergence conditions
of such iterative methods applied to
transfer problems of polarized radiation
is still lacking.
The specific
stability requirements are problem-dependent and
often difficult to determine.
This paper gives a deeper analysis of these convergence
conditions, analyzing their dependence on different design elements,
such as the formal solver,
the spectral, angular, and spatial numerical grids,
and the damping factor.

Section~\ref{sec:iterative_methods} briefly introduces
iterative methods, exposes their convergence conditions,
and presents the most common stationary methods.
Section~\ref{sec:crd_problem} presents the benchmark analytical problem
employed in this paper and exposes its discretization and algebraic formulation
in the context of various iterative methods.
Section~\ref{sec:results} numerically analyzes
the impact of different problem elements on the iterative solutions,
focusing on the choice of the formal solver, on the various discrete grids,
and on damping parameters.
Finally, Section~\ref{sec:conclusions} provides remarks and conclusions,
which are also generalized to more complex problems.
\section{Iterative methods }\label{sec:iterative_methods}
We consider the linear system
\begin{equation}\label{linear_system}
 A\xx=\bb,
\end{equation}
where the nonsingular matrix $A\in\mathbb R^{N\times N}$ and the vector $\bb \in \mathbb R^N$ are given, and the solution
$\xx  \in \mathbb R^N$ is to be found.
In absence of rounding errors, direct methods give the exact solution to the linear system~\eqref{linear_system} by a finite sequence of steps. They provide the solution
\begin{equation*}
  \xx=A^{-1}\bb,
\end{equation*}
using, for example, a suitable factorization of $A$, such as
Gaussian elimination.
Due to their high computational complexity and storage costs,
direct methods are often prohibitive in practice, especially for large problems.
For this reason, iterative methods are usually preferred for the solution of large linear systems. These methods are convenient in terms of parallel implementation and therefore particularly suitable for today's high performance computing systems.

Iterative methods start with an initial guess $\xx^0\in\mathbb R^N$ and generate a sequence of improving approximate solutions $\xx^1,\xx^2,\dots,\xx^n,\xx^{n+1}\in\mathbb R^N$ given by
\begin{align*}
 \xx^1 &= \phi_0(\xx^0;A,\bb),\\
 \xx^2 &= \phi_1(\xx^0,\xx^1;A,\bb),\\
 &\;\;\vdots\\
 \xx^{n+1} &= \phi_n(\xx^0,\xx^1,\dots,\xx^n;A,\bb).
\end{align*}
A one-step iterative method computes the approximate solution $\xx^{n+1}$ by considering only the previous approximate solution $\xx^n$. Hence, the transition from $\xx^n$ to $\xx^{n+1}$ does not rely on the previous history of the iterative process.
Moreover, a stationary iterative method generates the sequence of approximate solutions
by the recursive application of the same operator $\phi$. Therefore,
stationary one-step iterative methods can be expressed in the form
\begin{equation*}
 \xx^{n+1}=\phi(\xx^n;A,\bb).
\end{equation*}
If the operator $\phi$ is also linear,
the linear stationary one-step iterative method can be expressed in the form
\begin{equation}\label{linear_iter}
 \xx^{n+1}=G\xx^n+\mathbf{g},
\end{equation}
where neither the iteration matrix $G\in \mathbb R^{N\times N}$ nor the vector $\mathbf{g}$ depend upon the iteration $n$.
Hence, a linear stationary one-step iterative method starts from an initial guess $\xx^0$
and repeatedly invokes the recursive formula~\eqref{linear_iter}
until a chosen termination condition matches (cf. Section~\ref{sec:term_cond}).

In order to study and design linear stationary one-step iterative methods,
it is convenient to consider the linear system~\eqref{linear_system} in the form
\begin{equation}\label{linear_system2}
 \xx=G\xx+\mathbf{g}.
\end{equation}
We recall that multistep iterative methods have been also used in the radiative transfer context.
Relaying on multiple previous iterates, multistep iterative methods can accelerate the convergence of one-step iterations.
In this regard, we mention Ng \citep[e.g.,][]{olson1986}, ORTHOMIN \citep[e.g.,][]{klein1989}, and state-of-the-art Krylov 
methods. In particular, Krylov acceleration strategies are the subject of the second part of this paper series \citep{benedusi2021}.
\subsection{Convergence conditions}
An iterative method is convergent if the approximate solution $\xx^n$ converges to the exact solution $\xx$ as the iteration process progresses, that is, if $\xx^n\to \xx$ for $n\to\infty$.
This implies that the error, defined as
\begin{equation*}
    \mathbf{e}^n=\xx^n-\xx,
\end{equation*}
must approach zero as $n$ increases.
Subtracting Equation~\eqref{linear_system2} from Equation~\eqref{linear_iter}, one obtains
\begin{equation*}
 \mathbf{e}^{n+1}=G\mathbf{e}^n,
\end{equation*}
indicating that the convergence of the iterative method depends
entirely on the iteration matrix $G$.
In fact, the iterative method~\eqref{linear_iter} is convergent for any initial guess $\xx^0$ if and only if
\begin{equation}\label{conv_cond}
 \rho(G)<1,
\end{equation}
where $\rho(G)$ is the spectral radius of $G$ \citep[see, e.g.,][]{hackbusch2016}. The quantity $\rho(G)$ is also known as the convergence rate, because it describes the speed of convergence of an iterative process.
When employing an iterative method, a rigorous convergence analysis based on the condition~\eqref{conv_cond}
is advisable. However, since an analytical expression of $\rho(G)$ is often hard to obtain in practice, heuristic-based analyses are commonly employed to study how the convergence of iterative methods depends on problem and discretization parameters \citep{hackbusch2016}.
\subsection{Examples}\label{sec:examples}
Linear stationary one-step iterative methods 
for the solution of \eqref{linear_system}
usually arise from the splitting
\begin{equation}\label{splitting}
 A=P-R,
\end{equation}
with $P$ nonsingular. Rewriting \eqref{linear_system} in the form of \eqref{linear_system2} and introducing iteration indices as in \eqref{linear_iter}, one obtains
\begin{equation}\label{linear_iter1}
 \xx^{n+1}=P^{-1}R\xx^n+P^{-1}\bb,
\end{equation}
with $G=P^{-1}R=P^{-1}(P-A)=I\hspace{-0.1em}d-P^{-1}A$ and $\mathbf{g}=P^{-1}\bb$.
Crucially, $P$ has to be easier to invert then $A$,
while being a good approximation of it to ensure fast convergence.

Sometimes, it is convenient to express iterative methods in terms of the correction
\begin{equation*}
    \bm{\delta} \xx^n = \xx^{n+1}-\xx^n.
\end{equation*}
In the correction form, the iterative method~\eqref{linear_iter1} reads
\begin{equation*}
 \xx^{n+1}=(I\hspace{-0.1em}d-P^{-1}A)\xx^n+P^{-1}\bb = \xx^n + P^{-1}(\bb - A\xx^n),
\end{equation*}
or simply
\begin{equation}\label{linear_iter2}
 \xx^{n+1}=\xx^n + \bm\delta \xx^n,
\end{equation}
where the correction is given by $\bm\delta \xx^n = P^{-1}\rr^n$ and $\rr^n = \bb - A\xx^n$ is the iteration residual.

Many iterative methods are defined through the following decomposition
of the matrix $A$
\begin{equation*}
 A=D+L+U,
\end{equation*}
where $D$ is the diagonal of $A$, while
$L$ and $U$ are its strict lower and strict upper triangular parts, respectively.
Different choices of the matrix $P$ yield different iterative methods:
%
\begin{align*}
    P&=I\hspace{-0.1em}d, &&\;\;\text{ Richardson method},\\
    P&=D, &&\;\;\text{ Jacobi method},\\
    P&=\text{blockdiag}(A_1,\ldots,A_m), &&\;\;\text{ block-Jacobi method},\\
    P&=D+L \text{ or } P=D+U, &&\;\;\text{ Gauss-Seidel method},
\end{align*}
with $A_1,\ldots,A_m$ being $m$ square diagonal blocks of $A$.
All these choices of $P$ allow for a convenient,
and possibly parallel, computation of $P^{-1}$.
We note that the iterative method corresponding to $P=A$ always converges in one iteration, resulting in a direct method. 

In another perspective,
known as preconditioning,
all these methods correspond to a Richardson iteration applied to the 
preconditioned system
\begin{equation}\label{eq:prec_sys}
    P^{-1}A\xx=P^{-1}\bb,
\end{equation}
which is equivalent to the linear system~\eqref{linear_system}.
The nonsingular matrix $P$ is known as preconditioner and is usually chosen
to reduce the condition number of $P^{-1}A$ with respect to the one of $A$,
such that the preconditioned system is ``easier'' to solve
with respect to the original one.

Multiplying the preconditioner $P$ (or part of it) by a positive scalar, leads to damped iterative schemes, which are discussed in the following section.
\subsection{Damped iterative methods}\label{sec:damping}
The classical idea of damping (or relaxation) consists in scaling the correction $\bm{\delta} \xx^n$ in \eqref{linear_iter2} by a damping factor $\omega\in\mathbb R_{>0}$, namely
\begin{equation} \label{eq:prec_damped}
 \xx^{n+1}=\xx^n + \omega\bm{\delta} \xx^n = \xx^n + \omega P^{-1}\mathbf{r}^n.
\end{equation}
The term damped properly holds only for $0<\omega<1$, a.k.a. under-relaxation.
In fact, the choice $\omega=1$ yields the original method, while $\omega>1$ gives an extrapolated version, a.k.a. over-relaxation.
For simplicity, the term damped will be used for all choices of $\omega$.
Damped iterations can also be read as
\begin{equation*}
 \xx^{n+1}=\xx^n-\omega \xx^n+\omega \xx^n+\omega\bm{\delta} \xx^{n+1}=(1-\omega)\xx^n+\omega \tilde{\xx}^{n+1},
\end{equation*}
that is, the weighted mean of the old iterate $\xx^n$
and one step of the original method $\tilde{\xx}^{n+1} = \xx^n+\bm{\delta} \xx^n$. 

In order to guarantee the convergence of damped methods, it is possible to obtain bounds on $\omega$ as well as optimal choices in terms of convergence rate. In particular, the preconditioned iteration~\eqref{eq:prec_damped} is convergent if and only if 
\begin{equation}\label{eq:convergence_condition}
\omega < \frac{2\text{Re}(\lambda_i)}{|\lambda_i|^2}, \quad \text{for all} \quad i = 1,\ldots,N,
\end{equation}
$\lambda_i$ being the $i$th eigenvalue of the matrix $P^{-1}A$.
If $P^{-1}A$ has real positive eigenvalues,
the optimal convergence rate for the iterative methods presented in Section~\ref{sec:examples} is attained for
\begin{equation}\label{omega_opt}
 \omega_{\text{opt}}=\frac{2}{\lambda_{\min}+\lambda_{\max}},
\end{equation}
where $\lambda_{\min}$ and $\lambda_{\max}$ are the minimum and maximum eigenvalues
of $P^{-1}A$.
%
The SOR method is a damped variant of the Gauss-Seidel method, with $P = \omega^{-1}D+L$. If $A$ is consistently ordered, $G_{\text{Jac}}=I\hspace{-0.1em}d-D^{-1}A$ has real eigenvalues, and  $\rho(G_{\text{Jac}})<1$,
the optimal convergence rate for SOR is attained for 
\begin{equation}\label{omega_sor}
 \omega_{\text{sor}}=1+\left(\frac{\rho(G_{\text{Jac}})}{1+\sqrt{1-\rho(G_{\text{Jac}})^2}}\right)^2.
\end{equation}
%
A deeper and comprehensive analysis of damped stationary iterative methods is given, for instance,
by \citet{hageman1981,saad2003iterative,quarteroni2010numerical,hackbusch2016}. 
\subsection{Termination condition}\label{sec:term_cond}
Typically, iterative solution strategies terminate at the $n$th step if the norm  of the relative residual is smaller then a desired tolerance, that is, if $\| \rr^n\|_2/\|\bb\|_2<\text{tol}$. However, this stopping criteria becomes less and less reliable as the condition number of $A$, denoted with $\kappa(A)$, increases.
In fact, the following relation between the relative error and the relative residual holds:
\begin{equation*}
 \|\xx-\xx^n\|_2/\|\xx\|_2\leq\kappa(A)\|\rr^n\|_2/\|\bb\|_2.   
\end{equation*}
Since by construction $\kappa(P^{-1}A)<\kappa(A)$, the same termination condition with respect to the preconditioned system \eqref{eq:prec_sys}, namely, $\|P^{-1}\rr^n\|_2/\|P^{-1}\bb\|_2<\text{tol}$, gives a better approximation of the true relative error.
%

\section{Benchmark problem}\label{sec:crd_problem}
In this section, we present the continuous formulation of a linear benchmark transfer problem for polarized radiation. 
Secondly, we consider its discretization and derive its algebraic formulation.
Finally, we remark how the stationary iterative methods presented in Sections~\ref{sec:examples} and~\ref{sec:damping} are applied in this context.

The problem is formulated within the framework of the theoretical approach
described in~\citet{landi_deglinnocenti+landolfi2004}. Within this theory, a
scattering process is treated as a succession of independent absorption and reemission processes.
This is the so-called limit of complete frequency redistribution (CRD), in which
no frequency correlations between the incoming and the outgoing photons in a scattering process are taken into account.
We consider a two-level atom with an unpolarized lower level.
Since stimulated emission is completely negligible in solar applications,
it is not considered. 
For the sake of simplicity, 
the contribution of continuum processes is neglected, as well as that
of magnetic and bulk velocity fields.
We note, however, that the methodology presented in this section can also be applied to
a wider range of settings, such as the unpolarized case, two-term atomic models
including continuum contributions, atmospheric models with arbitrary magnetic and bulk velocity fields, 
and theoretical frameworks accounting for partial frequency redistribution (PRD) effects.

\subsection{Continuous problem}
%
The physical quantities entering the radiative transfer problem are, 
in general, functions of the spatial point $\rr$, the frequency $\nu$,
and the propagation direction ${\bf\Omega}$ of the radiation ray under consideration.
In the absence of polarization phenomena, the specific intensity and the atomic 
level population fully describe the radiation field and the atomic system, respectively. 
When polarization is taken into account,
a more detailed description is instead required.
The radiation field is fully described by the four Stokes parameters
$$I_i(\rr,{\bf\Omega},\nu),$$ 
with $i = 1,\ldots,4$, standing for Stokes $I$, $Q$, $U$, and $V$, respectively.
The $n$th energy level of the atomic system, with total angular momentum $J_n$, is fully described 
by the multipolar components of the density matrix, a.k.a. spherical statistical tensors,
$$\rho_{Q,n}^{K}(\rr),$$
with $K=0,\ldots,2J_n,$ and $Q=-K,\ldots,K$.
We recall that the spherical statistical tensors are in general complex
quantities, and that the 0-rank component, $\rho_{0,n}^0$, is proportional to the population 
of the $n$th level \citep[see][Chapter~3]{landi_deglinnocenti+landolfi2004}.
A two-level atomic system at position $\rr$ is thus described by
the following quantities
$$\rho_{Q,\ell}^{K}(\rr), \;\text{ and }\; \rho_{Q,u}^{K}(\rr),$$
with the subscripts $\ell$ and $u$ indicating
the lower and upper levels, respectively.
%
%
The assumption of unpolarized lower level\footnote{
This is strictly true when $J_\ell=0$. However, it is a good approximation 
when the lower level has a long lifetime (e.g., in case it is a ground or metastable
level) and the rate of elastic depolarizing collisions is high (i.e., when
the plasma density is high).}
implies that
\begin{equation*}
\rho^K_{Q,\ell}(\rr) = \rho^0_{0,\ell}(\rr) \delta_{K0} \delta_{Q0},
\end{equation*}
where $\delta_{ij}$ is the Kronecker delta.

In the absence of magnetic fields, the statistical equilibrium equations of a two-level atom 
with an unpolarized lower level
have the following analytical solution \citep[see][Chapter~14]{landi_deglinnocenti+landolfi2004}\footnote{
The statistical equilibrium equations additionally include
\begin{equation*}
\sqrt{2J_\ell + 1}\rho^0_{0,\ell}(\rr)+\sqrt{2J_u+1}\rho^0_{0,u}(\rr)=1,
\end{equation*}
which represents the conservation of the total number of atoms.}:
\begin{equation}
	\frac{ \rho^K_{Q,u}(\rr)}{\rho^0_{0,\ell}(\rr)} = 
	\sqrt{\frac{2J_u + 1}{2J_\ell + 1}}
	\frac{c^2}{2 h \nu_0^3}
	\left[ a_{J_u J_\ell}^{(K,Q)}(\rr)\bar{J}^K_{-Q}(\rr)
	+ \epsilon(\rr)W(\rr)\delta_{K0} 
	\delta_{Q0} \right],
	\label{Eq:SEE_rho_u}
\end{equation}
where $c$ and $h$ have their usual meaning of light speed and Planck constant, respectively,
$\epsilon$ is the so-called thermalization parameter, $W$ is the Planck function in the Wien limit at the 
line-center frequency $\nu_0$, and
\begin{equation*}
a_{J_u J_\ell}^{(K,Q)}(\rr)=(-1)^Q\frac{(1-\epsilon(\rr))
w^{(K)}_{J_u J_\ell}}{1 + \delta^{(K)}_u(\rr) (1-\epsilon(\rr))}.
\end{equation*}
The coefficient $w^{(K)}_{J_u J_\ell}$ is the
polarizability factor\footnote{This quantity is given by
\begin{equation*}
w^{(K)}_{J_u J_\ell} = (-1)^{1+J_u+J_\ell}\sqrt{3(2J_u+1)}
	\left\{ \begin{array}{c c c}
		1 & 1 & K \\
		J_u & J_u & J_\ell
	\end{array} \right\},
\end{equation*}
where the quantity in curly parentheses is the 6j-symbol
\citep[see][Chapter~2]{landi_deglinnocenti+landolfi2004}.},
$\delta^{(K)}_u$ describes the depolarizing rate of the upper level due 
to elastic collisions, and the frequency-averaged radiation field tensor is given by
\begin{equation}
	\bar{J}^K_Q(\mathbf{r})=
	\int {\rm d} \nu\,\phi(\mathbf{r},\nu)
	\frac{1}{4 \pi} \oint \mathrm{d} \mathbf{\Omega}
	\sum_{i=1}^4\mathcal{T}_{Q,i}^K(\mathbf{\Omega})
	I_i(\mathbf{r},\mathbf{\Omega},\nu),
\label{J_quad}
\end{equation}
where $\mathcal{T}^K_{Q,i}(\mathbf{\Omega})$ is the polarization tensor
\citep[see][Chapter~5]{landi_deglinnocenti+landolfi2004}
and $\phi$ is the absorption profile.
By definition, the maximum rank of the
radiation field tensor $\bar{J}^K_Q$ is $K=2$
\citep[see][Chapter~5]{landi_deglinnocenti+landolfi2004}. 
Consequently, the only nonzero 
spherical statistical tensors of the upper level 
$\rho^K_{Q,u}$ are those with $K\le2$.
Exploiting the conjugation properties of the spherical statistical tensors 
\citep[see][Chapter~3]{landi_deglinnocenti+landolfi2004}, at each position $\rr$, the considered 
two-level atomic system with an unpolarized lower level is thus described by 9 real quantities.

The propagation of the Stokes parameters
at frequency $\nu$ along $\mathbf{\Omega}$ is described by the
following differential equation
\begin{equation*}
	\frac{\mathrm{d}}{\mathrm{d}s}
	I_i(\rr,\vec{\Omega},\nu) = - \sum_{j=1}^4 K_{ij}(\rr,\vec{\Omega},\nu)
	I_j(\rr,\vec{\Omega},\nu) + 
	\varepsilon_i(\rr,\vec{\Omega},\nu),
\end{equation*}
where $s$ is the spatial coordinate along the direction $\bf{\Omega}$,
$\varepsilon_i$ is the emission coefficient of the $i$th Stokes parameter,
and the entries $K_{ij}$ form the $4\times4$ propagation matrix.

In the absence of a magnetic field and lower level polarization,
the propagation matrix is diagonal with diagonal element $\eta$.
The transfer equations for the four Stokes parameters
are consequently decoupled and read
\begin{equation}\label{RTE_delo}
	\frac{\mathrm{d}}{\mathrm{d}s}
	I_i(\rr,\vec{\Omega},\nu) = - \eta(\rr,\nu)
	\left[I_i(\rr,\vec{\Omega},\nu) - S_{\!i}(\rr,\vec{\Omega}) 
	\right].
\end{equation}
We note that $\eta$ does not depend on $\vec{\Omega}$ because 
bulk velocity fields are neglected, while the source function $S_{\!i}$ does not depend on $\nu$ 
because of the CRD assumption.
%
Considering dipole scattering and neglecting the continuum contribution,
the source function for the $i$th Stokes parameter is 
given by \citep[see][Chapter~14]{landi_deglinnocenti+landolfi2004}
\begin{equation}
	S_{\!i}(\rr,\vec{\Omega})= \frac{\varepsilon_i(\rr,\vec{\Omega},\nu)}{\eta(\rr,\nu)}
	=\sum_{KQ} w^{(K)}_{J_u J_\ell}\mathcal{T}^K_{Q,i}(\vec{\Omega})
	\sigma^K_Q(\rr),
	\label{mod_emission}
\end{equation}
%
with
\begin{equation}
	\sigma^K_Q(\rr) = \frac{2 h \nu_0^3}{c^2}
	\sqrt{\frac{2J_\ell + 1}{2J_u + 1}}
	\frac{ \rho^K_{Q,u}(\rr)}{\rho^0_{0,\ell}(\rr)}.
	\label{Eq:SKQ}
\end{equation}
Using \eqref{Eq:SEE_rho_u}, the quantities $\sigma^K_Q$, which 
can be interpreted as the multipolar components of the source function, are given by
\begin{equation}\label{Eq:see_SKQ}
	\sigma^K_Q(\rr) = a_{J_u J_\ell}^{(K,Q)}(\rr) \bar{J}^K_{-Q}(\rr) + \epsilon(\rr) W(\rr) \delta_{K0} \delta_{Q0}.
\end{equation}
\subsection{Discretization and algebraic formulation}
In order to replace the continuous problem by a system of algebraic equations,
we discretize the continuous variables $\nu$, $\rr$, and $\bf\Omega$ using, respectively, grids with $N_{\nu}$, $N_s$ and $N_\Omega$ nodes.
The quantities of the problem are now evaluated at the nodes at
$\{\rr_k\}_{k=1}^{N_s}$, $\{\mathbf{\Omega}_m\}_{m=1}^{N_\Omega}$, and $\{\nu_p\}_{p=1}^{N_{\nu}}$ only.

For $k=1,\ldots,N_s$, the quantity $\sigma^K_Q(\rr_k)$ is computed through the discrete versions of \eqref{Eq:see_SKQ} and \eqref{J_quad}, namely,
\begin{equation}\label{skq_crd}
 \sigma^K_Q(\rr_k)=\sum_{i=1}^4\sum_{m=1}^{N_\Omega}\sum_{p=1}^{N_\nu} J_{KQ,i}(\rr_k,{\bf\Omega}_m,\nu_p)I_i(\rr_k,{\bf\Omega}_m,\nu_p)+ c_Q^K(\rr_k),
\end{equation}
where $J_{KQ,i}(\rr_k,{\bf\Omega}_m,\nu_p)$ depends 
on the choice of spectral and angular quadratures used in~\eqref{J_quad} (cf.~\eqref{J_KQ_coeff}), while 
$c_Q^K(\rr_k)=\epsilon(\rr_k)W(\rr_k)\delta_{K0}\delta_{Q0}$.

Similarly, the discrete version of \eqref{mod_emission} reads
\begin{equation}\label{source_matrix_crd}
 S_{\!i}(\rr_k,{\bf\Omega}_m)=\sum_{KQ}T_{i,KQ}({\bf\Omega}_m)\sigma^K_Q(\rr_k),
\end{equation}
with $T_{i,KQ}({\bf\Omega}_m)=w^{(K)}_{J_u J_\ell} \mathcal{T}^K_{Q,i}({\bf\Omega}_m)$.
Finally, provided initial conditions,
the transfer equation~\eqref{RTE_delo} can be solved along each propagation direction 
${\bf\Omega}_m$ and at each frequency $\nu_p$ to provide the Stokes
component $I_i(\rr_k,{\bf\Omega}_m,\nu_p)$ for all $k$. This numerical step
can be expressed as
\begin{equation}\label{formal_solution_crd}
 I_i(\rr_k,{\bf\Omega}_m,\nu_p)=\sum_{l=1}^{N_s}\Lambda_i(\rr_k,\rr_l,{\bf\Omega}_m,\nu_p)S_{\!i}({\rr_l,\bf\Omega}_m)+t_i(\rr_k,{\bf\Omega}_m,\nu_p),
\end{equation}
where $t_i(\rr_k,{\bf\Omega}_m,\nu_p)$ represents the radiation transmitted from the boundaries and the coefficients $\Lambda_i(\rr_k,\rr_l,{\bf\Omega}_m,\nu_p)$ encode the numerical method used to propagate the $i$th Stokes parameter along the ray $\mathbf{\Omega}_m$ (cf. \eqref{Lam_coeff}). For notational simplicity, the explicit dependence on variables and indexes is omitted,
and Equations~\eqref{skq_crd}--\eqref{formal_solution_crd}
are expressed, respectively, in the following compact matrix form 
\begin{align}
\pmb{\sigma}&=J\mathbf{I}+\mathbf{c},&&\;\;\text{ with }
J\in\mathbb R^{9 N_s\times4 N_s N_\Omega N_\nu   },&&\label{matricial_form_1}\\
\mathbf{S}&=T\pmb{\sigma},&&\;\;\text{ with }
T\in\mathbb R^{4 N_s N_\Omega \times9 N_s},&&\label{matricial_form_2}\\
\mathbf{I}&=\Lambda\mathbf{S}+\mathbf{t},&&\;\;\text{ with }
\Lambda\in\mathbb R^{4  N_s N_\Omega N_\nu \times4 N_s N_\Omega },&&\label{matricial_form_3}
\end{align}
where the vector $\mathbf{I}\in\mathbb R^{4 N_s  N_\Omega N_\nu}$ collects the discretized Stokes parameters, 
$\mathbf{S}\in\mathbb R^{4 N_s N_\Omega }$ the discretized source function, and
$\pmb{\sigma}\in\mathbb R^{9 N_s}$
its discretized multipolar components.
The explicit expressions of vectors and matrices appearing in \eqref{matricial_form_1}--\eqref{matricial_form_3} are given in Appendices~\ref{app:Q}--\ref{app:Lambda} for a one-dimensional (1D) spatial grid.
In general, the entries of the matrix $J$ depend on the type of numerical integration used in Equation~\eqref{J_quad},
while the entries of $T$ depend on the coefficients of Equation~\eqref{mod_emission}.
Moreover, the entries of $\Lambda$ depend
on the numerical method used to solve the transfer Equation~\eqref{RTE_delo} (a.k.a. formal solver), on the spatial grid and, eventually, on the numerical conversion to the optical depth scale \citep[e.g.,][]{janett2018a}.

By choosing $\pmb{\sigma}$ as the unknown vector,
Equations~\eqref{matricial_form_1}--\eqref{matricial_form_3} can then be combined into a single discrete problem, namely,
\begin{equation}\label{crd_linear_system}
 (I\hspace{-0.1em}d-J\Lambda T)\pmb{\sigma}=J\mathbf{t}+\mathbf{c},
\end{equation}
with $I\hspace{-0.1em}d-J\Lambda T$ being a square matrix of size $9N_s$.

The operator $\Lambda$ and the vector $\mathbf{t}$ depend on the 
absorption coefficient $\eta$,
which linearly depends on $\rho^0_{0,\ell}$, which in turn enters the definition of the 
unknown $\pmb{\sigma}$. The discrete problem~\eqref{crd_linear_system} is consequently nonlinear.
However, the nonlinear problem~\eqref{crd_linear_system} becomes linear
if we further assume either that
%
(i) the transfer equation~\eqref{RTE_delo} 
is solved by introducing the optical depth scale defined by
\begin{equation*}
\mathrm{d} \tau(\nu) = -\eta(\rr,\nu)\mathrm{d}s,
\end{equation*}
and the optical depth grid is fixed a priori, or that
%
(ii) the density matrix element $\rho^0_{0,\ell}$ (or equivalently the lower level atomic
population) is fixed a priori and, consequently, the absorption coefficient $\eta$ becomes
a constant of the problem.
%
We note that in both cases
the operator $\Lambda$ and the vector $\mathbf{t}$
become independent of the unknown $\pmb{\sigma}$.

In the literature, the so-called $\Lambda$-operator
usually describes the mapping of the discretized
source function vector $\mathbf{S}$
either to the specific intensity vector $\mathbf{I}$ or to the corresponding mean radiation field vector \citep[e.g.,][]{olson1986,rybicki1991,paletou1995}.
This operator was introduced to highlight the linear nature
of this mapping inside the whole nonlinear radiative transfer problem.
The traditional $\Lambda$-operator corresponds either to~\eqref{matricial_form_3}
or to the combined action of \eqref{matricial_form_3} and the integral~\eqref{J_quad}.

We note that the linearized system~\eqref{crd_linear_system} has the same form as~\eqref{linear_system} with
$A=I\hspace{-0.1em}d-J\Lambda T$, $\xx=\pmb{\sigma}$, and
$\mathbf{b}=J\mathbf{t}+\mathbf{c}$.
The matrix $A$ can be assembled by constructing $J$, $\Lambda$, and $T$, computing the products $J\Lambda T$, and
finally calculating $I\hspace{-0.1em}d-J\Lambda T$.
Alternatively, the matrix $J\Lambda T$ can be assembled column-by-column,
constructing the $j$th column by applying \eqref{source_matrix_crd}--\eqref{formal_solution_crd}--\eqref{skq_crd} (setting $c_q^K$ and $t_i$ to zero) to a point-like multipolar component of the
source function $\pmb{\sigma}_i=\delta_{ij}$, for $i,j = 1,\ldots,9N_s$.

Applying the iterative methods exposed in Section~\ref{sec:examples} to Equation~\eqref{crd_linear_system}, the following iteration matrix is obtained
%
\begin{equation}\label{eq:G_matrix}
G = I\hspace{-0.1em}d- P^{-1}(I\hspace{-0.1em}d-J\Lambda T),
\end{equation}
$P$ being an arbitrary preconditioner.
Accordingly, the $n$th iteration in the correction form reads
\begin{equation*}
 \pmb{\sigma}^{n+1}=\pmb{\sigma}^n+\bm{\delta}\pmb{\sigma}^n,
\end{equation*}
where
\begin{equation*}
 \bm{\delta}\pmb{\sigma}^n= P^{-1}\left(J\Lambda T \pmb{\sigma}^n-\pmb{\sigma}^n
 +J\mathbf{t}+\mathbf{c}\right).
\end{equation*}
Sections~\eqref{subsec:richardson_crd}--\eqref{subsec:GS} describe
stationary iterative methods in the context of radiative transfer problems.
\subsection{Richardson method}\label{subsec:richardson_crd}
The form of the linear system~\eqref{crd_linear_system} suggests the operator splitting~\eqref{splitting} with
$P=I\hspace{-0.1em}d$ and $R=J\Lambda T$.
The iterative method based on this splitting, that is,
\begin{equation}\label{lambda_iteration_crd}
 \pmb{\sigma}^{n+1}=J\Lambda T \pmb{\sigma}^n+J\mathbf{t}+\mathbf{c},
\end{equation}
is an example of the Richardson method,
which is commonly known as $\Lambda$-iteration
in the context of linear radiative transfer problems.
Equation~\eqref{lambda_iteration_crd} describes 
the following iterative process:
(i) given an initial estimate of
$\pmb{\sigma}$,
calculate $\mathbf{S}$ at each discrete position and direction
through~\eqref{mod_emission};
 (ii) compute the numerical solution of the transfer equation~\eqref{RTE_delo}
to obtain $\mathbf I$ at each discrete position,
frequency, and direction;
 (iii) integrate $\mathbf I$ at each position according to~\eqref{J_quad}
to obtain the frequency averaged radiation field tensor, which is used
to update $\pmb{\sigma}$ using~\eqref{Eq:see_SKQ}.

The damped Richardson method is simply obtained 
by multiplying the correction $\bm{\delta}\pmb{\sigma}^n$ of the Richardson method by a
damping factor $\omega$.
\subsection{Jacobi method}\label{subsec:jacobi_crd}
Considering Equation~\eqref{crd_linear_system}, an usual operator splitting~\eqref{splitting} is given by choosing the preconditioner
$P$ to be diagonal, with the same diagonal of $I\hspace{-0.1em}d-J\Lambda T$, resulting in a minimal cost for the application and computation of $P^{-1}$, possibly in parallel. This splitting yields the Jacobi method, while its damped version is obtained by multiplying the correction $\bm{\delta}\pmb{\sigma}^n$ by a damping factor $\omega$.

\citet{trujillo_bueno1999} applied the Jacobi method
when dealing with CRD scattering line polarization linear problems.
In this context, the Jacobi method is sometimes termed as ALI method or local ALI method \citep[see][]{hubeny2003accelerated}.
However, an increasing problem size results in a lower impact of the Jacobi acceleration.
For instance, when 
PRD effects are included,
the size of the resulting linear system scales with $N_s N_\nu$ under the angle-averaged approximation
 \citep[see, e.g.,][]{alsinaballester2017},
while it scales with $N_s N_\nu N_\Omega$ in the general angle-dependent treatment \citep[see, e.g.,][]{janett2021b}.
In both cases, the impact of the Jacobi acceleration is barely perceptible.
\subsection{Block-Jacobi method}
In radiative transfer problems,
linear systems often have a natural block structure,
which offers different possibilities in defining the blocks.
Appendices~\ref{app:Q}--\ref{app:Lambda} explicitly illustrate
the structure of the matrices that build up the linear system~\eqref{crd_linear_system} for the 1D case.
Considering~\eqref{crd_linear_system}, 
the preconditioner 
$P$ can be chosen as the block-diagonal of $I\hspace{-0.1em}d-J\Lambda T$.
This choice yields the block-Jacobi method, while its damped version
is simply obtained by multiplying the correction $\bm{\delta}\pmb{\sigma}^n$ by a damping factor $\omega$.
We remark that, within this method, the computation of $P^{-1}$ requires solving the multiple smaller systems
resulting from the diagonal blocks.

\citet{trujillo_bueno1999} first applied the block-Jacobi method
to the CRD radiative transfer problem, choosing
square blocks that account for the coupling of
the different
multipolar components of the source function.
Moreover, the block-Jacobi method is a common choice
when dealing with PRD radiative transfer problems
under the angle-averaged approximation
\citep[e.g.,][]{belluzzi2014,alsinaballester2017,sampoorna2017,alsinaballester2018}.
In this case, the preconditioner is usually built by square blocks of size $N_\nu$
that account for the coupling of all the frequencies \citep[see the frequency-by-frequency method of][]{paletou1995}.
In such applications, the block-Jacobi method is referred to as ALI, Jacobi, or Jacobi-based method. \citet{janett2021b} first applied the damped block-Jacobi method
to the same problem.
\subsection{Gauss-Seidel method} \label{subsec:GS}
Another choice of the operator splitting~\eqref{splitting} is given by choosing the preconditioner $P$ as the lower (or upper) triangular part of $I\hspace{-0.1em}d-J\Lambda T$.
This splitting yields the Gauss-Seidel method.
The SOR method is simply obtained 
by modifying the correction $\bm{\delta}\pmb{\sigma}^n$ with a
suitable damping factor $\omega$ as explained in Section~\ref{sec:damping}.
\citet{trujillo_bueno1999} first proposed and applied
the Gauss-Seidel and SOR methods
to CRD scattering line polarization linear problems,
while \citet{sampoorna2010b} generalized these methods to the PRD case.
\citet{trujillo_bueno1995} confirmed with
demonstrative results that the optimal parameter $\omega_{\text{sor}}$ given by \eqref{omega_sor} is indeed effective in the context of numerical radiative transfer. However, Gauss-Seidel and SOR methods are not common in radiative transfer applications \citep{lambert2016},
mainly because they have limited and nontrivial parallelization capabilities
\citep[see, e.g., the red-black Gauss-Seidel algorithm][]{saad2003iterative}.
\section{Numerical analysis}\label{sec:results}
It is worth mentioning that realistic radiative transfer problems can be substantially more complex
than the one considered here. However,
we present a suitable benchmark to better understand
how the convergence of various iterative methods depends
on different design elements,
such as the choice of the formal solver, the discretization of the problem,
or the use of damping factors.
The conclusions could be then generalized to
more complex problems.

In the absence of magnetic and bulk velocity fields,
1D plane-parallel atmospheres satisfy cylindrical symmetry with respect to the vertical.
Due to the symmetry of the problem, the only nonzero
radiation field tensor components are $\bar{J}^0_0$ and $\bar{J}^2_0$ and,
from~\eqref{Eq:see_SKQ}, the only nonzero 
multipolar components of the source function are $\sigma^0_0$ and $\sigma^2_0$.
Setting the angle $\gamma=0$ in
the explicit expression of the polarization tensor 
$\mathcal{T}^2_{0,i}$ entering~\eqref{mod_emission},
the only nonzero source functions 
are $S_1$ and $S_2$ (i.e., $S_{\!I}$ and $S_{\!Q}$).
Consequently, the only nonzero Stokes parameters are $I_1$ and $I_2$ (i.e., $I$ and $Q$).
Moreover, the angular dependence of the problem variables
is fully described by the inclination $\theta\in[0,\pi]$ with
respect to the vertical, or equivalently by $\mu=\cos(\theta)$.
The atmosphere is assumed to be homogeneous and isothermal.
The thermalization parameter 
entering~\eqref{Eq:see_SKQ} is chosen as $\epsilon=10^{-4}$, while the damping constant entering 
the absorption profile $\phi$ as $a=10^{-3}$.
A two-level atom with $J_u=1$ and $J_\ell=0$ is considered and 
the depolarizing effect of elastic collisions is neglected.
All the problem parameters are summarized below
\begin{align*}
    & J_u  = 1,\quad J_l = 0, \quad w^{(0)}_{10} = w^{(2)}_{10} = 1, \\
    & \delta_u^{(K)}=0, \quad a_{10}^{(0,0)} = a_{10}^{(2,0)} = 1 - \epsilon,  \\
    & W = 1, \quad \epsilon=10^{-4}, \quad a=10^{-3}, \\
    & \mathcal{T}^0_{0,1}(\mu)=1, \quad \mathcal{T}^0_{0,2}(\mu)=0,\\
    & \mathcal{T}^2_{0,1}(\mu)=\sqrt{2}(3\mu^2-1)/4, \\
    & \mathcal{T}^2_{0,2}(\mu)=\sqrt{2}(3\mu^2-3)/4.
\end{align*}
%
In 1D geometries, the spatial vector $\rr$ can be replaced by the scalar height coordinate z.
Unless otherwise stated, hereafter, the atmosphere is 
discretized along the vertical direction through a grid in frequency-integrated optical depth ($\tau$).
In particular, a logarithmically spaced grid given by
\begin{equation*}
10^{-5} = \tau_1 < \tau_2 < \cdots < \tau_{N_s} = 10^4 
\end{equation*}
is considered in the following sections.
The spectral line is sampled with $N_\nu$ frequency nodes
equally spaced in the reduced frequency interval $[-5,5]$, and $N_\Omega$ Gauss-Legendre nodes are used to discretize $\mu\in[-1,1].$

For the sake of clarity, a limited number of formal solvers are analyzed:
the first-order accurate implicit Euler method, the second-order accurate DELO-linear and DELOPAR methods, and the third-order accurate DELO-parabolic method\footnote{We remark that
the DELOPAR method effectively provides third-order accuracy
in the case of a diagonal propagation matrix.}.
However, the numerical analysis performed here
can be unconditionally applied to any formal solver.
Further details on the specific properties of the aforementioned
formal solvers are given by \citet{janett2017a,janett2017b}.
Similarly to \citet{trujillo_bueno1999}, the block Jacobi preconditioner
is used with a block size of two that corresponds
to the $\sigma^0_0$ and $\sigma^2_0$ components local coupling.

\begin{figure}
\centering
 \includegraphics[width=0.49\textwidth]{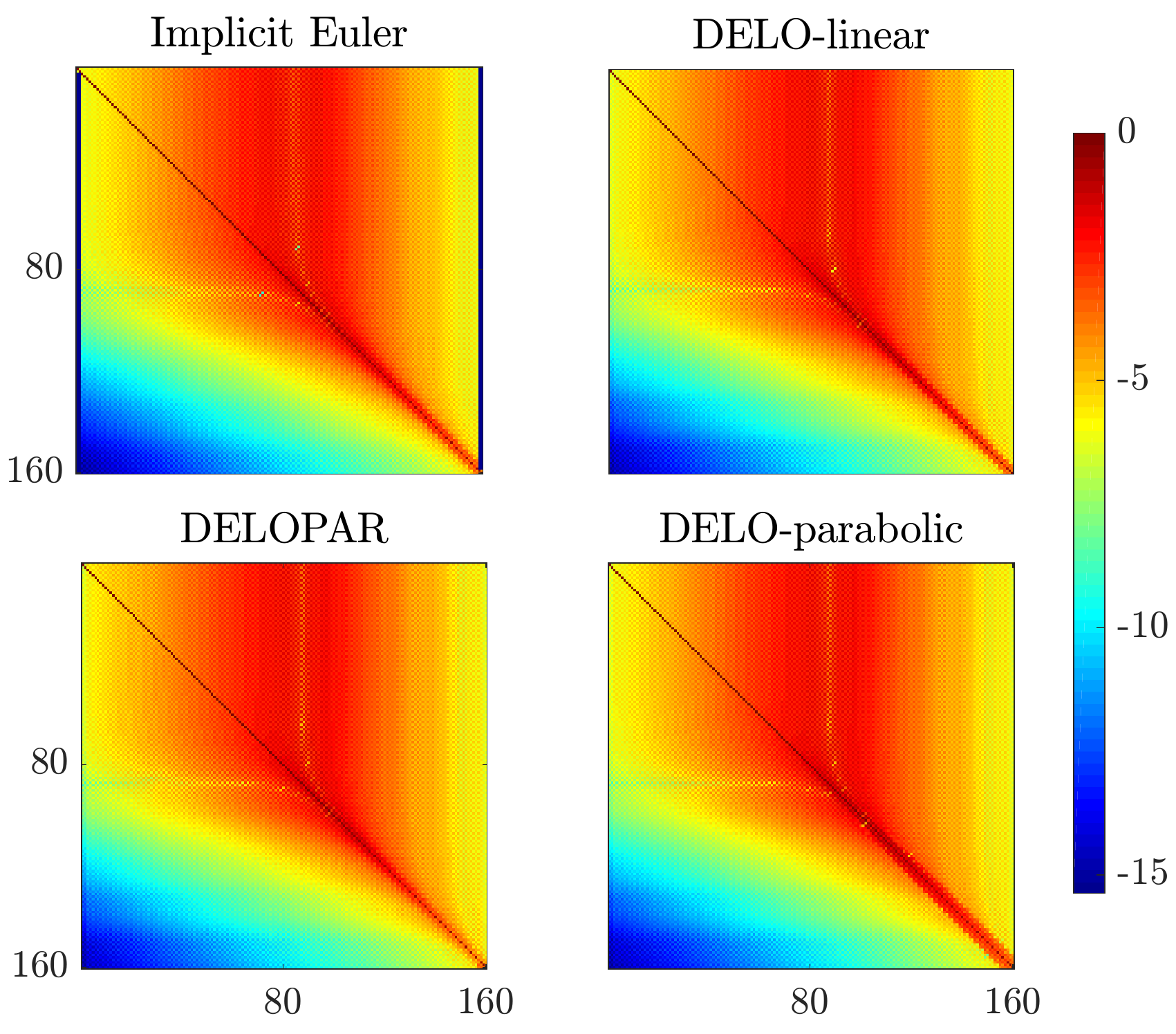}
 \caption{Values of $\log_{10}|Id-J\Lambda T|_{ij}$ for $i,j=1,\ldots,2N_s$ and multiple formal solvers, with $N_s=80$, $N_\Omega=21$, and $N_\nu=20$. }
\label{fig:matrix}
\end{figure}

In the following sections, we take advantage of the algebraic formulation of the transfer problem \eqref{crd_linear_system} to study the convergence of iterative methods presented in Section~\ref{sec:iterative_methods}. In particular, we focus on the spectral radius of the iteration matrix $\rho(G)$,
with $G$ defined in \eqref{eq:G_matrix}, for various preconditioners $P$ and formal solvers. We remark that the
line thermal contributions (encoded in $\mathbf{c}$) and boundary conditions (encoded in $\mathbf{t}$) enter only in the right-hand side of \eqref{crd_linear_system} and, consequently, they do not affect the results of this study. Figure~\ref{fig:matrix} displays the magnitude of the entries of the matrix $I\hspace{-0.1em}d-J\Lambda T$ for various formal solvers. Informally speaking, given the solution vector (cf. Appendix~\ref{app:Q})
\begin{equation*}
\pmb{\sigma} = [\sigma^0_0(\tau_1), \sigma^2_0(\tau_1),\sigma^0_0(\tau_2), \sigma^2_0(\tau_2),...,\sigma^0_0(\tau_{N_s}) \sigma^2_0(\tau_{N_s})]^T,
\end{equation*}
a large value of the entry $|Id-J\Lambda T|_{ij}$, with $i,j\in\{1,...,2N_s\}$, indicates that $\pmb{\sigma}_i$ depends strongly on $\pmb{\sigma}_j$. The asymmetry of the matrices is mainly due to the usage of a nonuniform spatial grid. The matrices entries show that,
as expected, for $\tau\leq 1$ (i.e., $i<90$) the solution $\pmb{\sigma}$ depends mostly on a region in the neighborhood 
of $\tau=1$. As $\tau$ increases this effect becomes less evident and the solution depends predominantly on neighboring nodes.
%
%

\subsection{Impact of spectral and angular quadratures}
\begin{figure}[ht]
     \quad \includegraphics[width=0.42\textwidth]{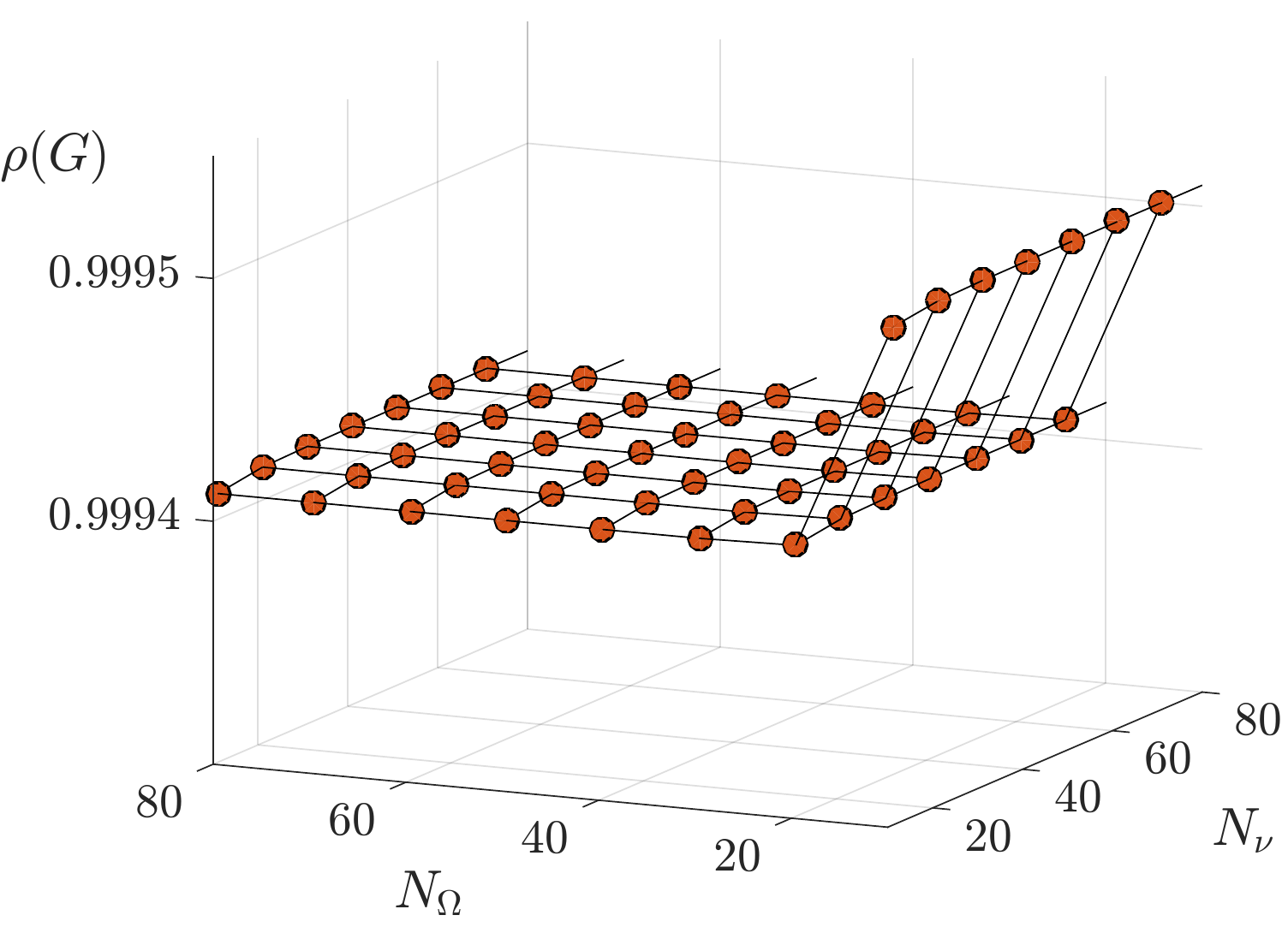}
    \caption{Spectral radius of the iteration matrix $\rho(G)$ for the DELO-linear formal solver and the Richardson iterative method, with $N_s=40$, varying $N_\Omega$ and $N_\nu$. The same behavior is observed for the other
    iterative methods and formal solvers analyzed, as well as for different values of $N_s$.}
    \label{fig:mu_nu_dependency}
\end{figure}

Figure~\ref{fig:mu_nu_dependency} illustrates that, once a minimal resolution is guaranteed, varying
the number of nodes in the spectral and angular grids $N_{\nu}$ and $N_{\Omega}$, respectively,
has a negligible effect on $\rho(G)$.
Therefore, the setting $N_{\nu}=21$ and $N_{\Omega}=20$ remains fixed in the following numerical analysis.
\subsection{Impact of the formal solver}
\begin{figure}
    \includegraphics[width=0.50\textwidth]{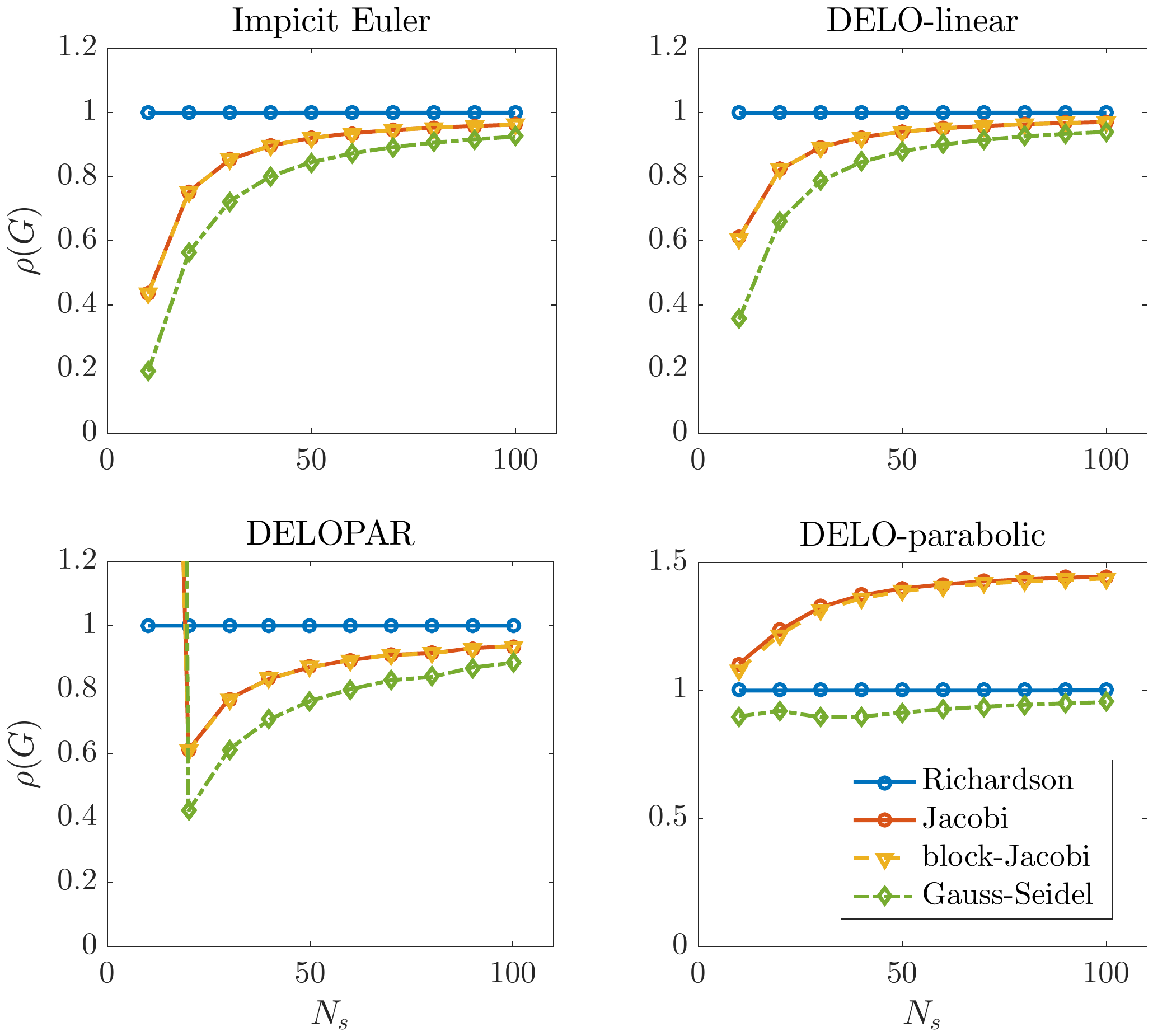}
    \caption{Spectral radius of the iteration matrix $\rho(G)$ for multiple formal solvers and undamped iterative methods with $N_\Omega=20$, $N_\nu=21$, and varying $N_s$.}
    \label{fig:conv_results}
\end{figure}

We now present the convergence rates of various undamped (i.e., with $\omega = 1$) iterative solvers, as a function of
the number of spatial nodes for various formal solvers.
The numerical results are summarized in Figure~\ref{fig:conv_results}. For the Richardson method (a.k.a. $\Lambda-$iteration) we observe that $\rho(G)\approx 1$, resulting in a very slow convergence. Indeed, this method is seldom used in practical applications.
Moreover, we generally observe a lower convergence rates as $N_s$ increases
for all formal solvers. 
Among the considered formal solvers, DELOPAR shows
the best convergence when combined with the 
Jacobi and Gauss-Seidel iterative methods.
For the same iterative methods, 
DELOPAR is not convergent for $N_s=10$, because in this case the matrix
$P^{-1}(I\hspace{-0.1em}d-J\Lambda T)$ has one negative eigenvalue, and thus \eqref{eq:convergence_condition} is not satisfied.
A particular case is given by the DELO-parabolic, for which the Jacobi and block-Jacobi iterative methods never converge. 
%
\subsection{Impact of the damping parameter}
\begin{figure}
\includegraphics[width=0.45\textwidth]{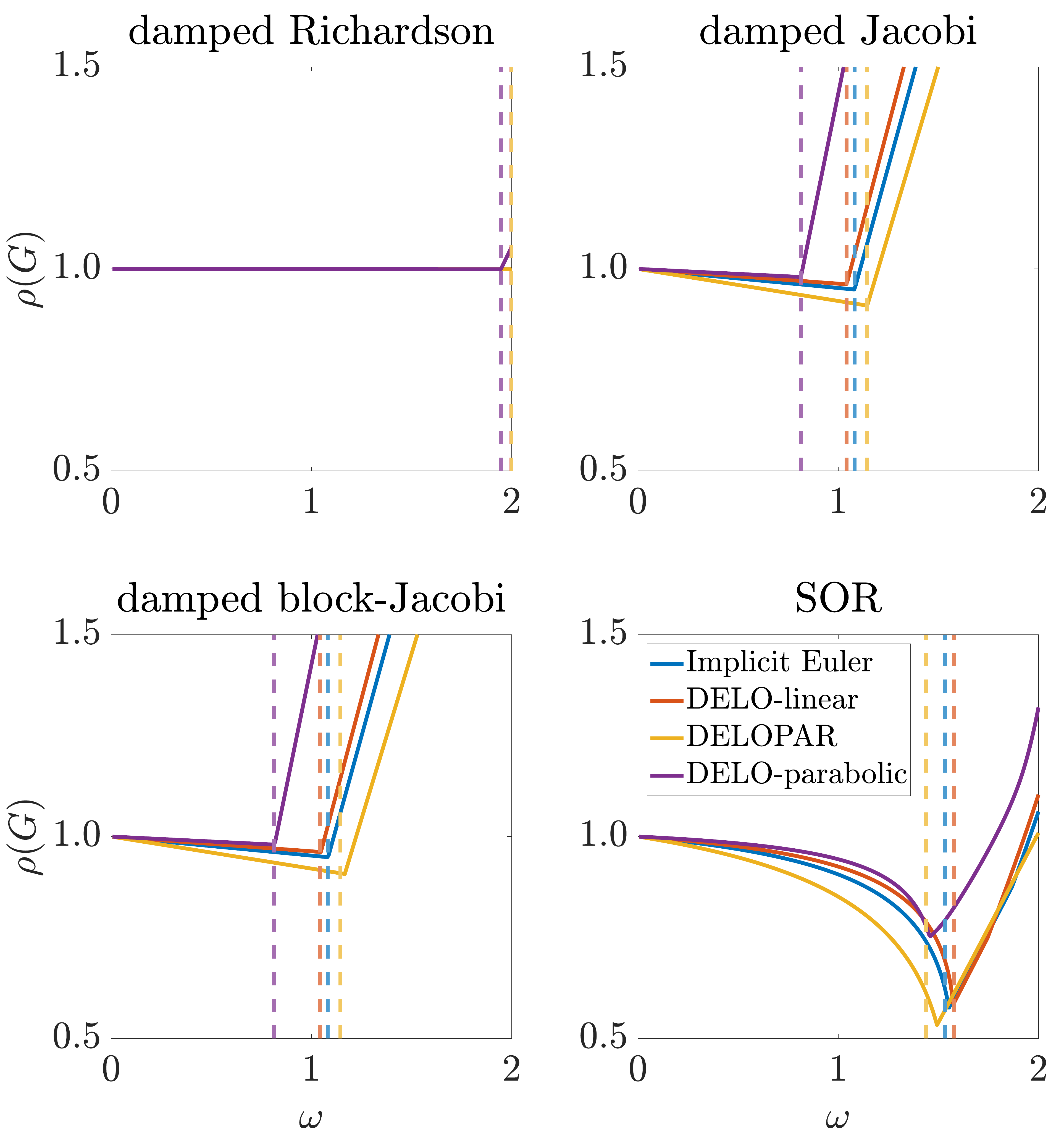}
\caption{Spectral radius of the iteration matrix $\rho(G)$ for multiple formal solvers and damped iterative methods with $N_\Omega=20$, $N_\nu=21$, $N_s=80$, and varying the damping parameter $\omega$.
Vertical dashed lines represent the optimal damping parameters defined in Section~\ref{sec:damping}.}
\label{fig:optimal_omega}
\end{figure}

We now discuss the impact of the damping parameter $\omega$
introduced in Section~\ref{sec:damping}. Figure~\ref{fig:optimal_omega} reports $\rho(G)$ for the damped methods as a function of the damping parameter $\omega$ and for multiple formal solvers.
The top left panel shows that the impact of the damping parameter
on the Richardson method is barely perceivable.
Similarly, top right and bottom left panels indicates that the impact of the damping parameter
on the Jacobi and block-Jacobi methods is not particularly pronounced.
In fact, the spectral radius provided by the optimal parameter does not differ substantially
from the value obtained with $\omega=1$.
However, a quite interesting result is revealed: the Jacobi and block-Jacobi methods (at $\omega=1$)
combined with the DELO-parabolic formal solver yield a spectral radius bigger than unity,
preventing convergence. However, the use of a suitable damping parameter ($\omega<0.81$) enforces stability,
guaranteeing convergence. An illustrative application of the damped block-Jacobi iterative method to
transfer of polarized radiation is given by \citet{janett2021b}.
The bottom right panel confirms that the impact of the damping parameter
in the SOR method is particularly effective.
Indeed, a suitable choice of the damping parameter significantly reduces
the spectral radius of the iteration matrix, increasing the convergence rate
of the SOR method.
This is consistent with the results of the previous investigations by
\citet{trujillo_bueno1995} and~\citet{trujillo_bueno1999}.
We notice that the optimal damping parameter cannot be calculated
with~\eqref{omega_sor} for the DELO-parabolic case, since $\kappa=\rho(G_{\text{Jac}})$ is larger then one.
Moreover, using the implicit Euler and DELOPAR formal solvers, the matrix $G_{\text{Jac}}$ has few complex eigenvalues and therefore
the estimate~\eqref{omega_sor} is not sharp.

\begin{figure}
    \includegraphics[width=0.49\textwidth]{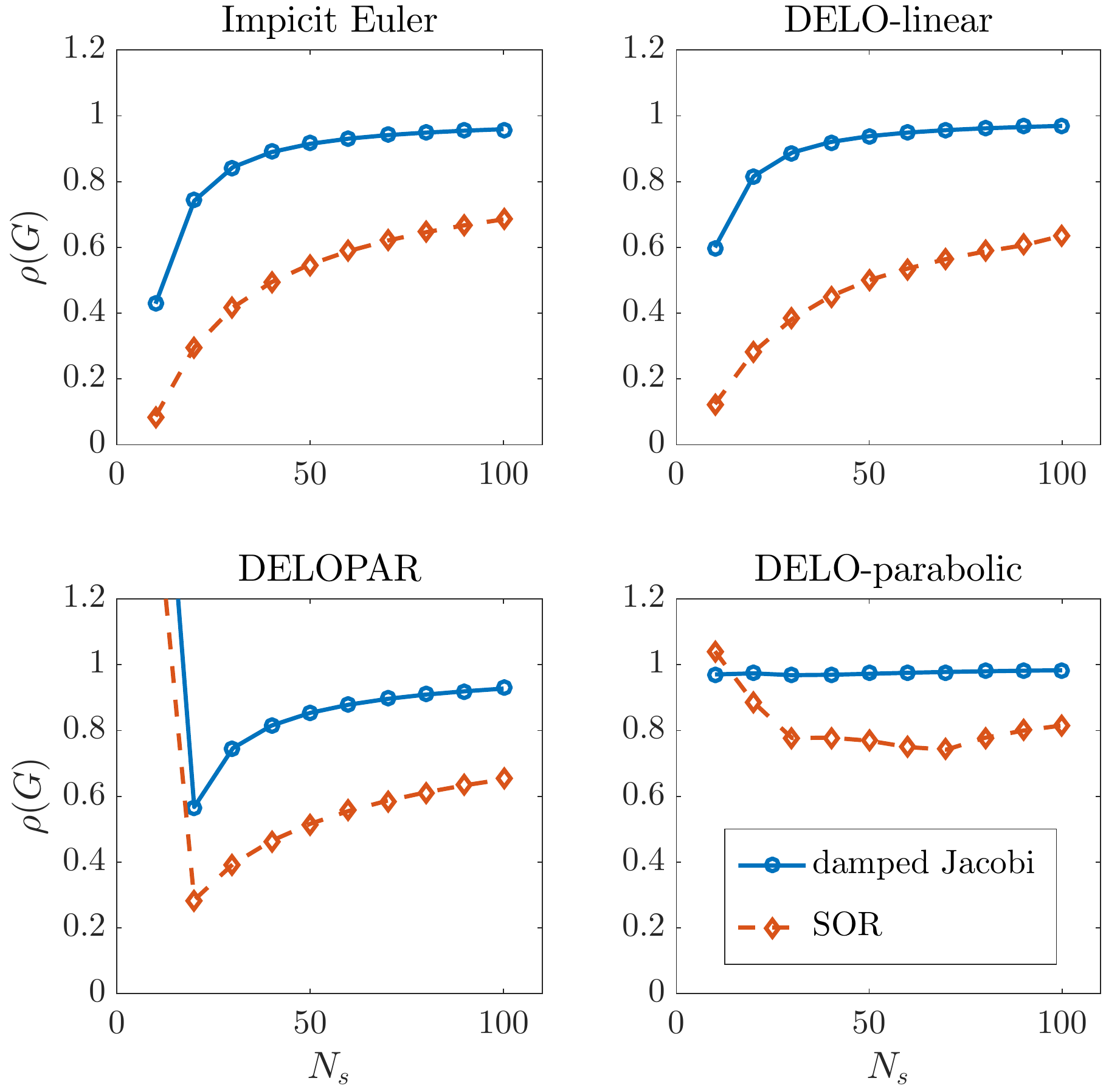}
\caption{Spectral radius of the iteration matrix $\rho(G)$ for multiple formal solvers and for the damped-Jacobi and SOR iterative methods with $N_\Omega=20$, $N_\nu=21$, and varying $N_s$.
For the damped-Jacobi and SOR methods, we used $\omega_{\text{opt}}$
from~\eqref{omega_opt} and $\omega_{\text{sor}}$ from~\eqref{omega_sor}, respectively.
Since~\eqref{omega_sor} cannot be applied if $\kappa<1$,
the $\omega_{\text{sor}}$ corresponding to DELOPAR is used
for the DELO-parabolic formal solver.}
\label{fig:conv_results_opt_omega}
\end{figure}

Figure~\ref{fig:conv_results_opt_omega} collects the convergence rates of the damped Jacobi and SOR methods, using the optimal damping parameters $\omega_{\text{opt}}$ and $\omega_{\text{sor}}$ defined in Section~\ref{sec:damping}. 
\subsection{Impact of the collisional destruction probability}
\begin{figure}
    \centering
    \includegraphics[width=0.49\textwidth]{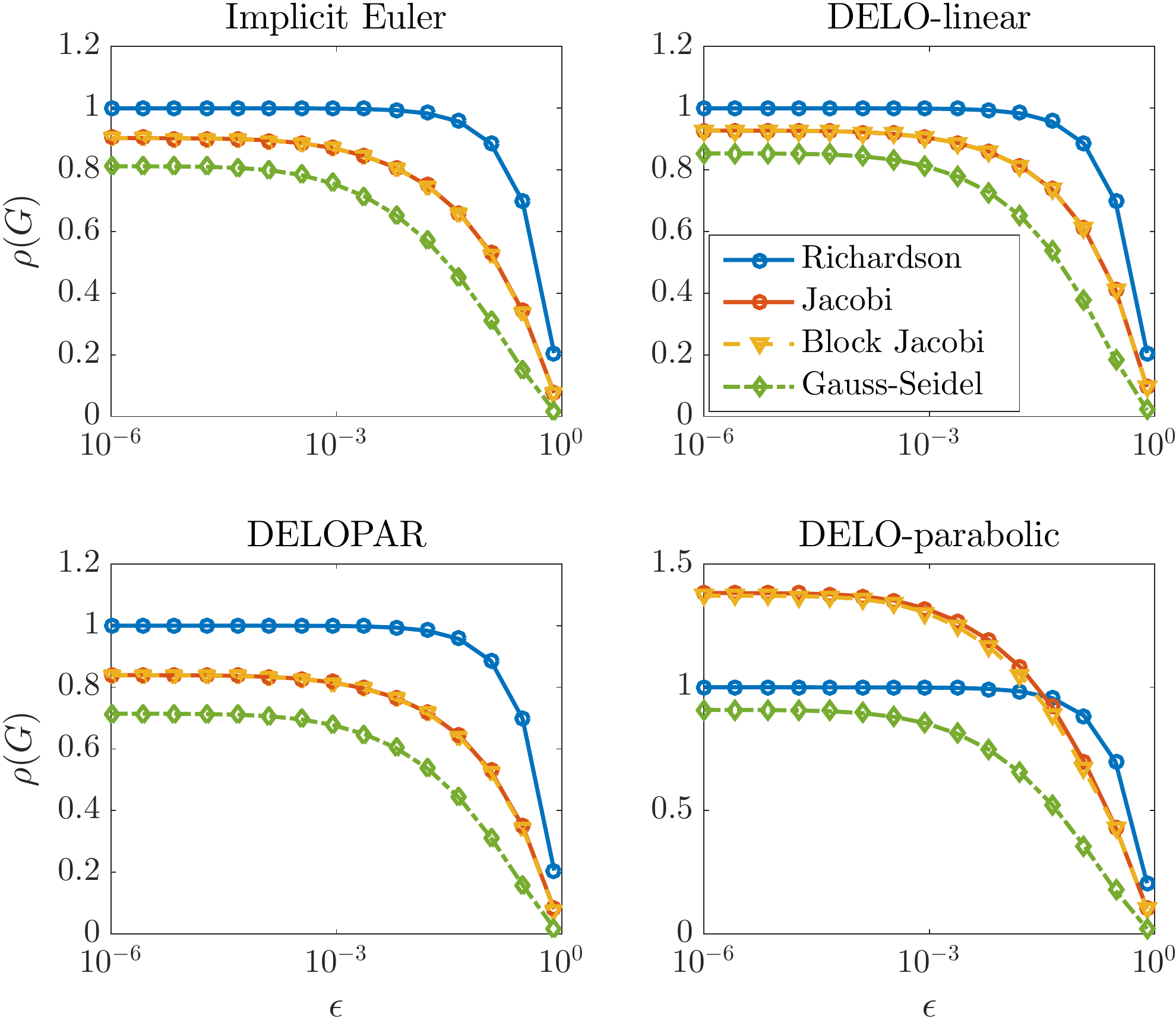}
\caption{Spectral radius of the iteration matrix $\rho(G)$ for multiple formal solvers and undamped iterative methods with $N_\Omega=20$, $N_\nu=21$, $N_s=40$, and varying the collisional destruction probability $\epsilon$.}
    \label{fig:eps_conv}
\end{figure}

Figure~\ref{fig:eps_conv} reports convergence rates for multiple $\epsilon \in[10^{-6},1)$, showing that a larger $\epsilon$ corresponds to faster convergence for all the iterative methods and formal solvers under investigation.
This result indicates that the numerical solution becomes easier when collision effects prevail. 

\section{Conclusions}\label{sec:conclusions}
This paper presents a fully algebraic formulation of a linear transfer problem for polarized radiation
allowing for a formal analysis that aims to better understand the convergence properties of stationary iterative methods.
In particular, we investigate how the convergence conditions of these methods depend on different problem and discretization settings, 
identifying cases where instability issues appear and exposing how to deal with them.

Numerical experiments confirm that the number of nodes in the spectral and angular grids
has a negligible effect on the spectral radius of iteration matrices of various methods and, hence, on their convergence properties.
By contrast, a larger number of spatial nodes leads to a reduction of convergence rates for all iterative methods and formal solvers
under investigation.

In general, the use of damping parameters can both enforce stability and increase the convergence rate.
Additionally, the optimal parameters described in Section~\ref{sec:damping} are indeed effective, especially in the SOR case.
In the particular case, where the Jacobi and block-Jacobi methods are combined with the DELO-parabolic formal solver, 
the spectral radius of the iteration matrix becomes larger than unity.
Hence, unless a suitable damping parameter is used, this combination is not convergent.

We remark that in this paper, we considered a specific benchmark linear radiative transfer problem.
However, the presented methodology can also be
applied to 
a wider range of settings, such as the unpolarized case, two-term atomic models 
including continuum contributions, 3D atmospheric models with arbitrary magnetic 
and bulk velocity fields, and theoretical frameworks accounting for PRD effects. 

Moreover, the algebraic formulation of the transfer problem
is used here exclusively as a practical tool for the convergence analysis. 
In the setting we have considered, the whole transfer problem is encoded in a dense linear system of size $2N_s$.
Different choices of the unknown vector ($\mathbf{I}$, $\pmb{\sigma}$, or $\mathbf{S}$) result in systems with different size, 
algebraic properties, and sparsity patterns.

However, this algebraic formulation paves the way to the application of advanced solution methods for linear (or nonlinear) systems, arising from radiative transfer applications.
In particular, many state-of-the-art parallel numerical techniques can be employed, such as Krylov methods \citep[see, e.g.,][]{lambert2015new}, 
parallel preconditioners
and multigrid acceleration \citep{hackbusch2016,saad2003iterative}.
Stationary iterative methods can also be applied as smoothers for multigrid techniques. 
These methods can also be applied in parallel, using available numerical libraries, which are tailored for high performance computing.
In particular, Krylov methods are the subject of study of the second part of this paper series \citep{benedusi2021}.

\begin{acknowledgements}
Special thanks are extended to E. Alsina Ballester, N. Guerreiro, and T. Simpson for particularly enriching comments on this work.
The authors gratefully acknowledge the Swiss National Science Foundation (SNSF) for financial support through grant CRSII5$\_180238$. Rolf Krause acknowledges the funding from the European High-Performance Computing Joint Undertaking (JU) under grant agreement No 955701 (project TIME-X). The JU receives support from the European Union's Horizon 2020 research and innovation programme and Belgium, France, Germany, Switzerland. 
\end{acknowledgements}

\bibliographystyle{aa}
\bibliography{bibfile}

\onecolumn

\appendix

\section{Matrix $J$}\label{app:Q}
The vector $\mathbf{I} \in \mathbb{R}^{4N_s N_\Omega N_\nu}$ collects
all the discrete values of the Stokes parameters with the order
defined by the tensor notation $\mathbf{I}\in\mathbb{R}^{N_s}\times \mathbb{R}^4\times \mathbb{R}^{N_\Omega}\times \mathbb{R}^{N_\nu}$, namely,
\begin{align}
\mathbf{I} = & [
I_1(\rr_1,\mathbf\Omega_1,\nu_1), 
I_1(\rr_1,\mathbf\Omega_1,\nu_2),\ldots,
I_1(\rr_1,\mathbf\Omega_1,\nu_{N_\nu}),
I_1(\rr_1,\mathbf\Omega_2,\nu_1),\ldots,
I_1(\rr_1,\mathbf\Omega_2,\nu_{N_\nu}),\ldots,
I_1(\rr_1,\mathbf\Omega_{N_\Omega},\nu_{N_\nu}),
I_2(\rr_1,\mathbf\Omega_1,\nu_1),\notag\\
 & \qquad \ldots,
 I_2(\rr_1,\mathbf\Omega_{N_\Omega},\nu_{N_\nu}),\ldots,
 I_4(\rr_1,\mathbf\Omega_{N_\Omega},\nu_{N_\nu}),
 I_1(\rr_2,\mathbf\Omega_1,\nu_1),\ldots,
 I_1(\rr_2,\mathbf\Omega_{N_\Omega},\nu_{N_\nu}),\ldots,
 I_4(\rr_{N_s},\mathbf\Omega_{N_\Omega},\nu_{N_\nu}]^T. \label{order_I}
 \end{align}
We also define the vector $\mathbf{I}_k\in\mathbb R^{4N_\Omega N_\nu}$ given by
\begin{equation*}
    \mathbf{I}_k=\left[I_1(\rr_k,\mathbf\Omega_1,\nu_1),\ldots,I_4(\rr_k,\mathbf\Omega_{N_\Omega},\nu_{N_\nu})\right]^T,
\end{equation*}
which collects the entries of $\mathbf{I}$ corresponding to the spatial point $\rr_k$, such that
$\mathbf{I}=\left[\mathbf{I}_1^T,\ldots,\mathbf{I}_{N_s}^T\right]^T$.

For $K=0,1,2$, $Q=-K,\ldots,K$ and $k=1,\ldots,N_s$ we express~\eqref{Eq:see_SKQ}
and~\eqref{skq_crd} as
\begin{equation}
\sigma^K_Q(\rr_k) = \sum_{i=1}^4\sum_{m=1}^{N_\Omega}\sum_{p=1}^{N_\nu} J_{KQ,i}(\rr_k,{\bf\Omega}_m,\nu_p)I_i(\rr_k,{\bf\Omega}_m,\nu_p)+ c_Q^K(\rr_k)= a_{J_u J_\ell}^{(K,Q)}(\rr_k)\mathbf{k}_{-Q}^K\mathbf{I}_k + c_Q^K(\rr_k),  \label{app_S_QK}
\end{equation}
where
\begin{equation}
J_{KQ,i}(\rr_k,{\bf\Omega}_m,\nu_p) = a_{J_u J_\ell}^{(K,Q)}(\rr_k)
\mathbf{k}^K_{-Q,N_\nu N_\Omega(i-1) + N_\nu(m-1) + p}.\label{J_KQ_coeff}
\end{equation}
The row vector $\mathbf{k}_{-Q}^K\in\mathbb R^{4N_\Omega N_\nu}$ collects the quadrature coefficients for the integral in \eqref{J_quad} and is given by the Kronecker product
\begin{equation}
    \mathbf{k}_Q^K = \frac{1}{4\pi}\bm{\psi}_Q^K\otimes\bm{\phi},
\end{equation}
where $\bm{\psi}^K_Q\in\mathbb{R}^{4N_\Omega}$ and $\bm{\phi}\in\mathbb{R}^{N_\nu}$
are given by
\begin{align*}
\bm{\psi}^K_{Q,(i-1)N_\Omega +m} & = \mathcal{T}^K_{Q,i}({\bf\Omega}_m) u_m, \text{ for } i=1,\ldots,4 \text{ and } m = 1,\ldots,N_\Omega,\\
\bm{\phi}_p & = \phi(\nu_p)v_p, \text{ for } p = 1,\ldots,N_\nu,
\end{align*}
and $u_m$ and $v_p$ are the quadrature weights of the grids
$\{\mathbf{\Omega}_m\}_{m=1}^{N_\Omega}$ and $\{\nu_p\}_{p=1}^{N_{\nu}}$,
respectively.

The matrix $J\in\mathbb R^{9 N_s\times4 N_s N_\Omega N_\nu}$ appearing in~\eqref{matricial_form_1}
is then given by 
$$ J =
\begin{bmatrix}
\bar{J}(\rr_1) & & &\\
& \bar{J}(\rr_2) & &\\
& & \ddots &  \\
& & & \bar{J}(\rr_{N_s}) \\
\end{bmatrix},
\qquad \text{ where } \qquad 
\bar{J}(\rr_k) = \begin{bmatrix}
a_{J_u J_\ell}^{(0,0)}(\rr_k)\mathbf{k}_0^0 \\
a_{J_u J_\ell}^{(1,-1)}(\rr_k)\mathbf{k}^1_1\\
a_{J_u J_\ell}^{(1,0)}(\rr_k)\mathbf{k}^1_0\\
\vdots \\
a_{J_u J_\ell}^{(2,1)}(\rr_k)\mathbf{k}^2_{-1} \\
a_{J_u J_\ell}^{(2,2)}(\rr_k)\mathbf{k}^2_{-2} \\
\end{bmatrix}.
$$
The matrix $\bar{J}(\rr_k)\in\mathbb R^{9\times4N_\Omega N_\nu}$,
which collects the coefficients  $a_{J_u J_\ell}^{(K,Q)}(\rr_k)\mathbf{k}_{-Q}^K$
is generally dense, but it also contains zero blocks, because $\mathcal{T}^K_{Q,i}({\bf\Omega}_m)$ is null for certain triples $i,K,Q$.

Accordingly to the structure of $J$, the vectors $\pmb{\sigma},\mathbf{c}\in\mathbb R^{9N_s}$ appearing in~\eqref{matricial_form_1} are given by 
\begin{align*}
& \mathbf{c} = \left[\tilde{\mathbf{c}}(\rr_1),\tilde{\mathbf{c}}(\rr_2),\ldots,\tilde{\mathbf{c}}(\rr_{N_s})\right]^T \quad
\text{ with } \quad \tilde{\mathbf{c}}(\rr_k) = \left[c_0^0(\rr_k),c_{-1}^1(\rr_k),c_0^1(\rr_k),\ldots,c_2^2(\rr_k)\right]\in\mathbb{ R}^9,\\
& \pmb{\sigma} = \left[\tilde{\pmb{\sigma}}(\rr_1),\tilde{\pmb{\sigma}}(\rr_2),\ldots,\tilde{\pmb{\sigma}}(\rr_{N_s})\right]^T \quad
\text{ with } \quad \tilde{\pmb{\sigma}}(\rr_k) = \left[\sigma^0_0(\rr_k),\sigma_{-1}^1(\rr_k),\sigma_0^1(\rr_k),\ldots,\sigma_2^2(\rr_k)\right]\in\mathbb{ R}^9.
\end{align*}
For completeness, we express the entries of the matrix $J$ in term of the coefficient appearing  in \eqref{skq_crd}, namely,
$$
J_{KQ,i}(\rr_k,{\bf\Omega}_m,\nu_p) = J_{ll'}, \quad \text{ with } \quad l=9(k-1) + K(K+1) + Q + 1 \quad \text{ and } \quad l'=4N_\Omega N_\nu(k-1) + N_\Omega N_\nu(i-1) + N_\nu(m-1) + p.
$$

\section{Matrix $T$}\label{app:S}
The vector $\mathbf{S} \in \mathbb{R}^{4N_s N_\Omega}$ collects
all the discrete values of the source function vector with the order
defined by the tensor notation $\mathbf{S}\in\mathbb{R}^{N_s}\times \mathbb{R}^4\times \mathbb{R}^{N_\Omega}$, namely,
\begin{equation*}
\mathbf{S} = [S_1(\rr_1,\mathbf{\Omega}_1),
S_1(\rr_1,\mathbf{\Omega}_2),\ldots,
S_1(\rr_1,\mathbf{\Omega}_{N_\Omega}),
S_2(\rr_1,\mathbf{\Omega}_1),\ldots,
S_2(\rr_1,\mathbf{\Omega}_{N_\Omega}),\ldots,
S_4(\rr_1,\mathbf{\Omega}_{N_\Omega}),
S_1(\rr_2,\mathbf{\Omega}_1),\ldots,
S_4(\rr_{N_s},\mathbf{\Omega}_{N_\Omega})
]^T.
\end{equation*}
We also define the auxiliary vector
$\mathbf{w}_i(\mathbf{\Omega})\in\mathbb R^9$ given by 
\begin{equation*}
\mathbf{w}_i(\mathbf{\Omega}) = \left[ 
w^{(0)}_{J_u J_\ell} \mathcal{T}^0_0(i,{\bf\Omega}),
w^{(1)}_{J_u J_\ell}\mathcal{T}^1_{-1}(i,{\bf\Omega}),
w^{(1)}_{J_u J_\ell}  \mathcal{T}^1_0(i,{\bf\Omega}),\ldots,
w^{(2)}_{J_u J_\ell}  \mathcal{T}^2_1(i,{\bf\Omega}),
w^{(2)}_{J_u J_\ell}  \mathcal{T}^2_2(i,{\bf\Omega})
\right].
\end{equation*}
The matrix $T\in\mathbb R^{4 N_s N_\Omega \times9 N_s}$ appearing in~\eqref{matricial_form_2}
is then given by
$$
T =
\begin{bmatrix}
W & & & \\
& W & & \\
& & \ddots &\\
& & & W \\
\end{bmatrix}, \qquad \text{ where } \qquad 
W = \begin{bmatrix}
\mathbf{w}_1({\bf\Omega}_1) \\
\mathbf{w}_1({\bf\Omega}_2) \\
\vdots \\
\mathbf{w}_1({\bf\Omega}_{N_\Omega}) \\
\mathbf{w}_2({\bf\Omega}_1) \\
\vdots \\
\mathbf{w}_4({\bf\Omega}_{N_\Omega}) \\
\end{bmatrix}.
$$
The matrix $W\in\mathbb R^{4N_\Omega\times9}$ is generally dense and
the matrix $T$ is also given by the Kronecker product 
$T =I\hspace{-0.1em}d_{N_s} \otimes W$, where $I\hspace{-0.1em}d_{N_s}$ is the identity matrix of size $N_s$.
%
\section{Matrix $\Lambda$ }\label{app:Lambda}
%
%
For the sake of clarity, the implicit Euler method is
used to integrate the transfer equation~\eqref{RTE_delo}.
However, the same analysis can be generalized to any formal solver.
For simplicity, we consider the 1D plane-parallel cylindrical symmetric atmospheric model  used in Section~\ref{sec:results}.

The angular dependence of the
problem is encoded by the scalar $\mu=\cos(\theta)$
with the corresponding discrete grid
\begin{equation*}
     -1 \leq \mu_1 < \mu_2 < \ldots < \mu_{N_\Omega -1} < \mu_{N_\Omega} \leq 1.
     \end{equation*}
The spatial dependence of the
problem is encoded either by the scalar $s$
with the corresponding discrete grid $\{s_k\}_{k=1}^{N_s}$
or by the scaled line-center frequency optical depth $\tau$
with the corresponding discrete grid $\{\tau_k\}_{k=1}^{N_s}$.
The relation between the two spatial variables is described by
$$\phi(\nu)\mathrm{d}\tau=-\eta(s,\nu)\mathrm{d}s$$
%
and the transfer equation~\eqref{RTE_delo} can then be expressed as
\begin{equation}\label{RTE_delo_opt}
	\frac{\mu}{\phi(\nu)}\frac{\mathrm{d}}{\mathrm{d}\tau}
	I_i(\tau,\mu,\nu) = I_i(\tau,\mu,\nu) - S_{\!i}(\tau,\mu).
\end{equation}
Applying the implicit Euler method to \eqref{RTE_delo_opt} for $i=1,\ldots,4$, we obtain, for incoming  and outgoing directions respectively
\begin{align}
    I_i(\tau_{k+1},\mu,\nu) & = \frac{I_i(\tau_{k},\mu,\nu) +\Delta \tau_k(\mu,\nu)S_{\!i}(\tau_{k+1},\mu)}{1+\Delta \tau_k(\mu,\nu)},\qquad 
    \text{ for } \quad \mu<0 \quad \text{ and } \quad  k = 1,\ldots,N_{s}-1,\label{ie_1}\\
    I_i(\tau_{k-1},\mu,\nu) & = \frac{I_i(\tau_k,\mu,\nu) +\Delta \tau_{k-1}(\mu,\nu)S_{\!i}(\tau_{k-1},\mu)}{1+\Delta \tau_{k-1}(\mu,\nu)},  \qquad
    \text{ for } \quad \mu>0 \quad \text{ and } \quad  k = N_s,N_{s}-1,\ldots,2, \label{ie_2}
\end{align}
where
\begin{equation*}
\Delta \tau_{k}(\mu,\nu)=\frac{\phi(\nu)}{|\mu|}|\tau_{k+1}-\tau_k|.
\end{equation*}
We notice that
different Stokes parameters, directions and frequencies 
are decoupled in the transfer equation~\eqref{RTE_delo} and, consequently,
in the discrete counterparts~\eqref{ie_1} and~\eqref{ie_2}. Providing the initial conditions $I_i(\tau_1,\mu,\nu)=I_i^{\text{in}}(\mu,\nu)$ for $\mu<0$ and $I_i(\tau_{N_s},\mu,\nu)=I_i^{\text{out}}(\mu,\nu)$ for $\mu>0$,
we recursively apply~\eqref{ie_1} and~\eqref{ie_2} and obtain 
\begin{align}
    I_i(\tau_k,\mu,\nu)  & = f_k(\mu,\nu) I_i^{\text{in}}(\mu,\nu) + \sum_{j=2}^k f_{k,j}(\mu,\nu)S_{\!i}(\tau_j,\mu) \qquad \text{ for } \quad \mu<0 \quad \text{ and } \quad  k = 2,\ldots,N_s,\label{ie_mono1}\\
    I_i(\tau_k,\mu,\nu)  & = h_k(\mu,\nu) I_i^{\text{out}}(\mu,\nu) + \sum_{j=k}^{N_s-1} h_{k,j}(\mu,\nu)S_{\!i}(\tau_j,\mu) \qquad  \text{ for } \quad \mu>0 \quad \text{ and } \quad  k = 1,\ldots,N_s-1,\label{ie_mono2}
\end{align}
with
\begin{align}
    f_{k,j}(\mu,\nu) & = \frac{\Delta \tau_{j-1}(\mu,\nu)}{\prod_{l=j}^{k}  1 + \Delta \tau_{l-1}(\mu,\nu)} \quad \text{and} \quad f_k(\mu,\nu) = \frac{1}{\prod_{l=1}^{k-1}  1 + \Delta \tau_{l}(\mu,\nu)},\label{ie_mono3}\\ 
    h_{k,j}(\mu,\nu) & = \frac{\Delta \tau_{j}(\mu,\nu)}{\prod_{l=k}^{j}  1 + \Delta \tau_{l}(\mu,\nu)} \quad \text{and} \quad h_k(\mu,\nu) = \frac{1}{\prod_{l=k}^{N_s-1}  1 + \Delta \tau_{l}(\mu,\nu)}.\label{ie_mono4}
\end{align}
Equation~\eqref{ie_mono1} for $i=1,\ldots,4$
can be arranged in a linear system
with a lower triangular matrix $F\in\mathbb R^{N_s\times N_s}$, namely,
$$ \begin{bmatrix}
I_i(\tau_1,\mu,\nu) \\
I_i(\tau_2,\mu,\nu)\\
\vdots \\
I_i(\tau_{N_s},\mu,\nu) \\
\end{bmatrix} = 
\begin{bmatrix}
0 & & &  \\\
& f_{2,2}(\mu,\nu) & &  \\
& \vdots & \ddots &  \\
 & f_{N_s,2}(\mu,\nu)   & \cdots   & f_{N_s,N_s}(\mu,\nu)\\
\end{bmatrix}
\begin{bmatrix}
S_{\!i}(\tau_1,\mu) \\
S_{\!i}(\tau_2,\mu) \\
\vdots \\
S_{\!i}(\tau_{N_s},\mu) \\
\end{bmatrix} + 
\begin{bmatrix}
1 \\
f_2(\mu,\nu) \\
\vdots \\
f_{N_s}(\mu,\nu) \\
\end{bmatrix}I_i^{\text{in}}(\mu,\nu)
\Longleftrightarrow  \tilde{\mathbf{I}}_i(\mu,\nu) =  F(\mu,\nu)\mathbf{S}_i(\mu) + \ff_i(\mu,\nu).$$
Analogously, Equation~\eqref{ie_mono2} for $i=1,\ldots,4$
can be arranged in a linear system
with an upper triangular matrix $H\in\mathbb R^{N_s\times N_s}$, namely,
$$ \begin{bmatrix}
I_i(\tau_1,\mu,\nu) \\
I_i(\tau_2,\mu,\nu)\\
\vdots \\
I_i(\tau_{N_s},\mu,\nu) \\
\end{bmatrix} = 
\begin{bmatrix}
h_{1,1}(\mu,\nu) & \cdots & \cdots h_{1,N_s- 1}(\mu,\nu) & \\
& \ddots &  \vdots &\\
 & & h_{N_s-1,N_s-1}(\mu,\nu) &\\
 & & & 0
\end{bmatrix}
\begin{bmatrix}
S_{\!i}(\tau_1,\mu) \\
S_{\!i}(\tau_2,\mu) \\
\vdots \\
S_{\!i}(\tau_{N_s},\mu) \\
\end{bmatrix} +
\begin{bmatrix}
h_1(\mu,\nu) \\
\vdots \\
h_{N_s-1}(\mu,\nu) \\
1 \\
\end{bmatrix}I_i^{\text{out}}(\mu,\nu)
\Longleftrightarrow \tilde{\mathbf{I}}_i(\mu,\nu) =  H(\mu,\nu)\mathbf{S}_i(\mu)+\mathbf{h}_i(\mu,\nu).$$
The two previous linear systems can then be combined into a single linear system of size $N_s\times N_s$ given by
\begin{equation}
\tilde{\mathbf{I}}_i(\mu,\nu) =  L(\mu,\nu)\mathbf{S}_i(\mu)+\mathbf{l}_i(\mu,\nu),
\end{equation}
%
with 
\begin{equation*}
  L(\mu,\nu) =
    \begin{cases}
      F(\mu,\nu) & \text{if $\mu < 0$,} \\
      H(\mu,\nu) & \text{if $\mu > 0$,}
    \end{cases}  \qquad \text{and } \qquad
      \mathbf{l}_i(\mu,\nu) =
    \begin{cases}
      \mathbf{f}_i(\mu,\nu) & \text{if $\mu < 0$,} \\
      \mathbf{h}_i(\mu,\nu) & \text{if $\mu > 0$.}
    \end{cases}
\end{equation*}
%
For $k,l=1,...,N_s$, $m=1,N_\Omega$ and $p=1,..,N_\nu$, the entries of the matrix $\Lambda\in\mathbb R^{4  N_s N_\Omega N_\nu \times4 N_s N_\Omega }$
appearing in~\eqref{matricial_form_3} and in~\eqref{formal_solution_crd}, are then given by 
\begin{align}
& \Lambda_{k'l'} =\Lambda_i(\tau_k,\tau_l,\mu_m,\nu_p)=\left(L(\mu_m,\nu_p)\right)_{k,l} \qquad \text{ with } &&k' = 4N_\Omega N_\nu(k-1) + N_\Omega N_\nu(i-1) + N_\nu(m-1) + p, \label{Lam_coeff}\\
& &&l'= 4N_\Omega (l-1) + N_\Omega (i-1) + m.\nonumber
\end{align}
%
%
Accordingly, the vector $\mathbf{t}$ appearing in~\eqref{matricial_form_3}
reads
\begin{equation*}
\mathbf{t} = \left[
t_1(\tau_1,\mu_1,\nu_1),\ldots,
t_4(\tau_1,\mu_{N_\Omega},\nu_{N_\nu}),
t_1(\tau_2,\mu_1,\nu_1),\ldots,
t_4(\tau_2,\mu_{N_\Omega},\nu_{N_\nu}),
t_1(\tau_3,\mu_1,\nu_1),\ldots,
t_4(\tau_{N_s},\mu_{N_\Omega},\nu_{N_\nu})
\right]^T,
\end{equation*}
with
\begin{equation*}
t_i(\tau_k,\mu_m,\nu_p) = \left(\mathbf{l}_i(\mu_m,\nu_p)\right)_k.
\end{equation*}
that is using the same ordering of \eqref{order_I}.

%

For the sake of clarity, we only considered the implicit Euler method for the explicit assembly of the matrix $\Lambda$.
However, this analysis can be unconditionally applied to any formal solver, such as exponential integrators or multistep methods.
In practice, it would be sufficient to modify~\eqref{ie_mono3} and \eqref{ie_mono4} according to the selected formal solver.

\end{document}